\documentstyle[11pt,epsf]{article}
\newcommand {\dfn} {\stackrel{\Delta} {=}}

\newcommand {\reals} {{\rm I\!R}}

\newcommand {\bx} {\mbox{\boldmath $x$}}

\newcommand {\bE} {\mbox{\boldmath $E$}}

\newcommand {\bX} {\mbox{\boldmath $X$}}

\newcommand{\calX}{{\cal X}}

\begin{document}
\thispagestyle{empty}
\title{Relations Between Redundancy Patterns of the Shannon Code and
Wave Diffraction Patterns of Partially Disordered Media
}
\author{Neri Merhav
}
\date{}
\maketitle

\begin{center}
Department of Electrical Engineering \\
Technion - Israel Institute of Technology \\
Haifa 32000, ISRAEL \\
\end{center}
\vspace{1.5\baselineskip}
\setlength{\baselineskip}{1.5\baselineskip}

\begin{abstract}
The average redundancy of the Shannon code, $R_n$, as a function of the block
length $n$, is known to exhibit two very different types of behavior, depending on the
rationality or irrationality of certain parameters of the source: 
It either converges to $1/2$ as $n$ grows without
bound, or it may have a non--vanishing, oscillatory, (quasi--) periodic pattern
around the value $1/2$ for all large $n$. In this paper, we make an attempt to
shed some insight into this erratic behavior of $R_n$, by drawing an analogy
with the realm of physics of wave propagation, in particular, the elementary
theory of scattering and diffraction.
It turns out that there are two types of
behavior of wave diffraction patterns formed by crystals, which are
correspondingly analogous to the two types of patterns of $R_n$. When the crystal is
perfect, the diffraction intensity spectrum exhibits very sharp peaks, a.k.a.\
Bragg peaks, at wavelengths of full constructive interference. These wavelengths 
correspond to the frequencies of the harmonic waves of the oscillatory mode of
$R_n$. On the other hand, when the crystal is imperfect and there is 
a considerable degree of disorder in its structure, the Bragg peaks disappear,
and the behavior of this mode is analogous to the one where $R_n$ is
convergent.\\

\noindent
{\bf Index Terms:} Lossless source coding, redundancy, Shannon code,
scattering, diffraction, Bragg peaks, disorder.

\end{abstract}

\section{Introduction}

The analysis of the average redundancy of lossless codes for data compression
schemes is a topic that attracted the attention of considerably many researchers
throughout the history of Information Theory (cf.\ e.g.,
\cite{CD92},\cite{Gallager78},\cite{Huffman52},\cite{JS95},\cite{Krichevskii68},\cite{LS97},\cite{Rissanen86},\cite{Savari98},\cite{SG97},\cite{Szp00}
and many references therein). 

In \cite{Szp00} Szpankowski has derived the
asymptotic behavior of the average redundancy $R_n$, as a function of the
block length $n$, for the Shannon code, the Huffman
code, and other codes, focusing primarily on the binary memoryless source,
parametrized by $p$ -- the probability of zero. His analysis revealed a
rather interesting behavior of $R_n$, especially in the cases of the
Shannon code and the Huffman code: When $\alpha\dfn \log_2[(1-p)/p]$
is irrational, then $R_n$ converges to a constant (which is $1/2$ for the Shannon code
and $3/2-1/\ln 2$ for the Huffman code) as $n\to \infty$. On the other hand,
when $\alpha$ is rational, $R_n$ has a non--vanishing oscillatory term of the
form $\left<\beta m_0 n\right>$, where $\beta\dfn-\log_2(1-p)$, $m_0$ is the
denominator of $\alpha=\ell_0/m_0$ in its representation as the ratio between
two integers whose greatest common divisor is $1$, and 
$\left<x\right>=x-\lfloor x\rfloor$ designates the fractional part of a real number $x$.
In several places in his paper, Szpankowski describes this 
behavior of $R_n$ as ``erratic'' and this qualifier
is, of course, understandable. 

Our purpose in this paper is to make an attempt to give some insight into this
erratic behavior of $R_n$ by drawing an analogy with the physics of wave diffraction.
From the theory of X--ray scattering (see, e.g.,
\cite[Chapter 2]{CL95},\cite{Welberry85}), it is known that if the object 
that causes the diffraction of an incident wave is a
perfect crystal, then the intensity profile of the scattered wave (as a function of
the wavelength or the wave number) exhibits very sharp peaks, known as
{\it Bragg peaks}, at wavelengths
that correspond to full coherence, where the optical distance differences to all scattering
elements (layers of the crystal)
are exactly integer multiples of the wavelength. This continues to be the case as long
as there is enough order in the medium such that all these distances 
are commensurable and therefore have a common
divisor (common unit of length), which can serve as the fundamental wavelength. In the realm of the
average redundancy analysis, this corresponds to the case where $\alpha$ is
rational and the fundamental frequency of the oscillatory term $\left<\beta
m_0n\right>$ of $R_n$ is intimately related to the fundamental wavelength at which
there is a Bragg peak. On the other hand, when the distances are
incommensurable, perfect 
coherence between all scattered waves is not achieved at any wavelength 
and therefore no Bragg peaks are observed. This is the case of strong
disorder, which in the lossless source coding problem, corresponds to the case
of $\alpha$ irrational, where $R_n$ is convergent.

More concretely, the analysis of the scattered wave intensity function is based on a very
simple model of disorder, which is due to Hendricks and Teller \cite{HT42}
(see also \cite{GL88}).
According to the Hendricks--Teller (HT) model, the distances between every two
consecutive layers in the solid are selected independently at random from a finite set of
two or more distances. In the simplest case, where there are only two possible
distances $d_0$ and $d_1$, with probabilities $p$ and $1-p$, this random
selection process is analogous to the memoryless binary source of the data
compression problem and the parameter $\alpha$ of this source plays a role
analogous to that of the ratio $d_1/d_0$. Thus, $\alpha$ irrational means that $d_0$ and $d_1$
are incommensurable, which is the case of strong disorder with no Bragg peaks
and no oscillations in $R_n$. On the other hand when $\alpha=d_1/d_0$ is
rational, we are in the (partially) ordered mode, as described above.

From the pure mathematical point of view, the analogy between the average
redundancy problem and the diffraction problem is rooted in that
at the heart of the analyzes of both problems,
there is one very simple mathematical fact in
common: Given a vector $(p_0,p_1,\ldots,p_{M-1})$
of non--negative reals summing to unity (probabilities) and a vector 
$(\alpha_1,\ldots,\alpha_{M-1})\in\reals^{M-1}$, the complex number
\begin{equation}
\label{mainpoint}
C_m=p_0+\sum_{j=1}^{M-1}p_je^{2\pi i m\alpha_j},~~~i=\sqrt{-1},~~m=1,2,3,\ldots
\end{equation}
has a modulus that obviously never exceeds unity, and $C_m=1$ 
(i.e., full coherence between all $M$ phasors) is attained for
some integer values of $m$ 
if and only if $\{\alpha_j\}$ are all rational. When this is the case, then 
$C_m=1$ for all values of $m$ which are integer multiples of $m_0$, the
first positive integer $m$ for which $m\alpha_j$ is integer for all $1\le j \le
M-1$ at the same time.\footnote{The previous paragraph refers to the special
case $M=2$.} The analogy between the Shannon code redundancy analysis and
the diffraction patterns under the HT model will center around
(\ref{mainpoint}) and its two types of behavior depending on the rationality
or irrationality of $\{\alpha_j\}$.

The remaining part of this short 
paper consists of two more main sections. For the sake
of completeness, in Section 2,
we summarize the main ingredients of the derivation in \cite{Szp00} (with a
few shortcuts), emphasizing the use of the simple mathematical fact
described in the previous paragraph. For reasons of simplicity, we focus on
the Shannon code and the derivation specializes on the memoryless case.
In Section 3, we bring the derivation of the diffraction
patterns of the HT model, with a focus on the analogy with Section 2. We then
describe in detail the mapping between the two problems under discussion.
Finally, in Section 4 we summarize and conclude, with a comments on a possible
extension to the Markov case.

\section{Average Redundancy of the Shannon Code}

Throughout the remaining part of this paper, we use capital letters to
designate random variables (e.g., $X_i$) and the corresponding lower
case letters to denote specific realizations (e.g., $x_i$). 

Consider a finite alphabet memoryless source $X_1,X_2,\ldots$ with alphabet
$\calX=\{0,1,2,\ldots,M-1\}$ and symbol probabilities
$\{p_0,p_1,\ldots,p_{M-1}\}$.
The Shannon code for lossless data compression assigns to every
source $n$--tuple $\bx=(x_1,x_2,\ldots,x_n)\in\calX^n$ a binary codeword of length
\begin{equation}
\ell(\bx)=\lceil-\log P(\bx)\rceil=\lceil-\log\prod_{t=1}^n p_{x_t}\rceil,
\end{equation}
where $\lceil u\rceil$ designates the smallest integer not smaller than $u$.
The average redundancy of the Shannon code is defined as
\begin{equation}
R_n=\bE\left\{\ell(\bX)\right\}-nH
\end{equation}
where
\begin{equation}
H=-\sum_{j=0}^{M-1}p_j\log p_j
\end{equation}
is the per--symbol entropy.
The derivation of the asymptotic expression
for $R_n$ in \cite{Szp00} can be presented (with a few slight shortcuts and modifications)
as follows. By using the Fourier series expansion of the function
$\left<u\right>$, according to 
\begin{equation}
\label{fourier}
\left<u\right>=\frac{1}{2}-\sum_{m\ne 0} a_m e^{2\pi i
mu},~~~~a_m=\frac{1}{2\pi i m},
\end{equation}
we have the following:
\begin{eqnarray}
\label{Rn}
R_n&=&\bE\{\lceil-\log P(\bX)\rceil+\log P(\bX)\}\nonumber\\
&=&1-\bE\{-\log P(\bX)-\lfloor-\log P(\bX)\rfloor\}\nonumber\\
&=&1-\bE\left<-\log P(\bX)\right>\nonumber\\
&=&1-\bE\left\{\frac{1}{2}-\sum_{m\ne 0}a_m \exp\left[-2\pi i m\log
P(\bX)\right]\right\}\nonumber\\
&=&\frac{1}{2}+\sum_{m\ne 0}a_m\bE\left\{\exp\left[-2\pi im\log
P(\bX)\right]\right\}\nonumber\\
&=&\frac{1}{2}+\sum_{m\ne 0}a_m\sum_{\bx\in\calX^n}\left(\prod_{t=1}^n
p_{x_t}\right)\cdot\exp\left[-2\pi im\sum_t\log p_{x_t}\right]\nonumber\\
&=&\frac{1}{2}+\sum_{m\ne 0}a_m\sum_{\bx\in\calX^n}\prod_{t=1}^n
\left(p_{x_t}\exp\left[-2\pi im\log p_{x_t}\right]\right)\nonumber\\
&=&\frac{1}{2}+\sum_{m\ne 0}a_m\prod_{t=1}^n\left(\sum_{x_t=0}^{M-1}
p_{x_t}\exp\left[-2\pi im\log p_{x_t}\right]\right)\nonumber\\
&=&\frac{1}{2}+\sum_{m\ne 0}a_m\left(\sum_{j=0}^{M-1}
p_j\exp\left[-2\pi im\log p_j\right]\right)^n\nonumber\\
&=&\frac{1}{2}+\sum_{m\ne 0}a_me^{-2\pi i mn\log p_0}\left[p_0+\sum_{j=1}^{M-1}
p_j\exp\left\{2\pi im\log(p_0/p_j)\right\}\right]^n.
\end{eqnarray}
Denoting $\alpha_j=\log(p_0/p_j)$, $j=1,2,\ldots,M-1$, the expression in the
square brackets is exactly $C_m$ as was defined in (\ref{mainpoint}). The
behavior of $R_n$ for large $n$ is then as follows. If $\{\alpha_j\}$ are not
all rational, then $|C_m| < 1$ for all $m$, and so, $\lim_{n\to\infty}
C_m^n=0$, which causes the entire summation over $m$ to vanish for large $n$.
In this case, $R_n\to 1/2$ as $n\to\infty$. On the other hand, if
$\{\alpha_j\}$ are all rational, then there exists an integer $m$ such that
$m\alpha_j$ are all integers. Let $m_0$ be the smallest positive integer with
this property. Then all other integers with the same property are integer
multiples of $m_0$.
Consequently, $\lim_{n\to\infty} C_m^n=1$ whenever $m$ is an
integer multiple of $m_0$ and $\lim_{n\to\infty} C_m^n=0$ otherwise. Thus, 
denoting $\beta=-\log p_0$, we now have for large $n$,
\begin{eqnarray}
\label{osc}
R_n&\approx& \frac{1}{2}+\sum_{k\ne 0}a_{km_0}e^{2\pi i km_0n\beta}\nonumber\\
&=& \frac{1}{2}+\frac{1}{m_0}\sum_{k\ne 0}a_{k}e^{2\pi i km_0n\beta}\nonumber\\
&=& \frac{1}{2}+\frac{1}{m_0}\left(\frac{1}{2}-\left<\beta m_0
n\right>\right),
\end{eqnarray}
where the second line holds since $a_m$ is inversely proportional to $m$
(see (\ref{fourier}) above) and in the third line we used again 
(\ref{fourier}) with $u=\beta m_0 n$. As can easily be seen from the second
line of (\ref{osc}), for large $n$, the sequence $R_n$ is harmonic with a
fundamental frequency $\omega_0=2\pi m_0\beta$. In other words, the Fourier
transform of $\{R_n\}$ contains Dirac delta functions at integer multiples of
$\omega_0$ (modulo $2\pi$). We will see later on that these spectral 
spikes are analogous to the Bragg peaks of the HT model.

At this point, a technical comment is in order. At first glance, it may seem
that the above approximate expression of $R_n$ is assymetric with respect to 
permutations of the alphabet, because $\beta$ was defined as $-\log p_0$ and the choice of the
symbol $x=0$ as having a special role in the last line of (\ref{Rn}) was
completely arbitrary (we could have chosen, of course, any other symbol $j$ as well).
However, note that $\left<\beta m_0 n\right>=\left<-m_0
n\log p_0\right>$ is identical to $\left<-m_0 n\log p_j\right>$ for all
$j=1,\ldots,M-1$ because in the rational case considered above, 
the numbers $\{-m_0n\log p_j\}_{j=0}^{M-1}$ differ from each other by
integers, and therefore their fractional parts are all the same. Thus, the
above expression of $R_n$ is, in fact, invariant to permutations of the
alphabet.

\section{Diffraction Patterns of the HT Model}

The simplest way to think of the HT model is as a one--dimensional model of an
alloy, which is characterized by a sequence of
mass points, positioned along the real line at random locations $Z_0,
Z_1,\ldots,Z_{n-1}$. The ensemble of the HT model is defined in terms of the
spacings $\Delta_j\dfn Z_j-Z_{j-1}$, $j=1,2,\ldots,n-1$, 
which are $n-1$ i.i.d.\ random variables taking on values in a
finite set $\{d_0,d_1,\ldots,d_{M-1}\}$ with probabilities
$p_0,p_1,\ldots,p_{M-1}$, respectively (thus, $Z_0,Z_1,\ldots$ is a random
walk). Each point $Z_i$ contributes a
scattered wave described by the phasor $e^{-iq Z_j}$, where in the
one--dimensional setting considered here, $q$ can be
understood as the wave number, that is, $q=2\pi/\lambda$, where $\lambda$ is
the wavelength. Assuming the same amplitudes at all points, 
the superposition of all these contributions is
then the sum $U(q)=\sum_j e^{-iq Z_j}$, which can be interpreted as the Fourier
transform of the function $u(z)=\sum_j\delta(z-Z_j)$. The overall intensity of
this superposition of waves is designated by the {\it structure function}
\cite[Chapter 2]{CL95}
\begin{equation}
I(q)=\bE\{|U(q)|^2\}=\bE\left\{\sum_{k,\ell}
e^{iq(Z_k-Z_\ell)}\right\}=\sum_{k,\ell}\bE\{e^{iq(Z_k-Z_\ell)}\},
\end{equation}
where the expectation is with respect to the random variables $\{Z_j\}$.

The derivation of $I(q)$ is fairly simple (see, e.g., \cite{GL88}) and it is
brought here for the sake of completeness. 
\begin{eqnarray}
\label{decomp}
I(q)&=& \sum_{k,\ell} \bE\{e^{iq(Z_k-Z_\ell)}\}\nonumber\\
&=& n+\sum_{k > \ell} \bE\{e^{iq(Z_k-Z_\ell)}\}+\sum_{k < \ell}
\bE\{e^{iq(Z_k-Z_\ell)}\}\nonumber\\
&=& n+\sum_{k > \ell} \bE\{e^{iq(Z_k-Z_\ell)}\}+\sum_{k > \ell}
\bE\{e^{-iq(Z_k-Z_\ell)}\}\nonumber\\
&\dfn& n+I_0(q)+I_0^*(q)
\end{eqnarray}
where $I_0(q)$ is defined as the second term of the third line and $I_0^*(q)$
is the complex conjugate of $I_0(q)$. Now,
\begin{eqnarray}
I_0(q)&=&\sum_{k > \ell} \bE\{e^{iq(Z_k-Z_\ell)}\}\nonumber\\
&=&\sum_{k > \ell}\bE\left\{\exp\left[iq\sum_{s=\ell+1}^k
\Delta_s\right]\right\}\nonumber\\
&=&\sum_{k > \ell}\bE\left\{\prod_{s=\ell+1}^k\exp\left[iq
\Delta_s\right]\right\}\nonumber\\
&=&\sum_{k > \ell}\left[\sum_{j=0}^{M-1}p_je^{iqd_j}\right]^{k-l}\nonumber\\
&=&\sum_{r=1}^{n-1}(n-r)[C(q)]^r,
\end{eqnarray}
where we have denoted
\begin{equation}
C(q)=\sum_{j=0}^{M-1}p_je^{iqd_j}.
\end{equation}
For $n$ large, whenever
$|C(q)| < 1$, the last expression is dominated by the term
$n\sum_{r=1}^\infty [C(q)]^r=nC(q)/[1-C(q)]$, which together with the two
other terms of (\ref{decomp}), yields
\begin{equation}
I(q)\approx
n\left(1+\frac{C(q)}{1-C(q)}+\frac{C^*(q)}{1-C^*(q)}\right)=n\cdot\frac{1-|C(q)|^2}{|1-C(q)|^2},
\end{equation}
or equivalently,
\begin{equation}
\hat{I}(q)=\lim_{n\to\infty}\frac{I(q)}{n}=\frac{1-|C(q)|^2}{|1-C(q)|^2}.
\end{equation}
If there are values of $q$ for which $|C(q)|=1$, yet $C(q)\ne 1$, then the
geometric series diverges at these points, but these are only points of removable 
discontinuity in $\hat{I}(q)$ because for every other point, arbitrarily close to
such a discontinuity point,
again $|C(q)|<1$, and the
geometric series converges. The real problematic points, if any, are those where
$C(q)=1$ if they exist. For $C(q)=1$, we have to re-derive the expression of
$I(q)$ separately, which is very simple as $I(q)$ is just the sum of $n^2$
$1$'s, namely, $I(q)=n^2$. In other words, the intensity scales quadratically
rather than linearly with $n$, which means that these are extremely high peaks in $I(q)$,
namely, the Bragg peaks.

For $C(q)$ to take the value $1$ for some $q$, the products $qd_j$ must
all be integer multiples of $2\pi$. Suppose that $q$ is such that $qd_0=2\pi m$ for some
integer $m$, i.e., $q=q_m\dfn 2\pi m/d_0$, in which case we shall denote
$C(q_m)$ by $C_m$, as before. In this case,
\begin{equation}
C_m=p_0+\sum_{j=1}^{M-1}p_j e^{2\pi i m d_j/d_0}.
\end{equation}
But this is again exactly the expression in (\ref{mainpoint}), this time with
$\alpha_j=d_j/d_0$, which as mentioned earlier, may assume the value 1, for
some integer values of $m$, if and only if
$\alpha_j=d_j/d_0$ are all rational, or equivalently, $d_0,d_1,\ldots,d_{M-1}$
are commensurable. When this is the case, then as before, there exists an
integer $m$
for which $md_j/d_0$ are all integers simultaneously. Analogously to the
derivation in Section 2, let $m_0$ be the smallest integer with this property.
Then, the Bragg peaks appear at wave-numbers $q_{km_0}$, $k=1,2,\ldots$, which
correspond to wavelengths $\lambda_0/k$, where $\lambda_0=d_0/m_0$.

The analogy between the two settings is now clear: The memoryless source 
of Section 2 is parallel to the random selection process in the HT model. 
The parameters $\alpha_j=\log (p_0/p_j)$ of the source are analogous to
distance ratios $d_j/d_0$, $j=1,2,\ldots,M-1$. Their rationality/irrationality
dictates the mode of behavior in both problems.
The integer parameter $m_0$ is
then defined in both settings in the very same way. The partially ordered mode
in the diffraction model is parallel to the oscillatory mode of $R_n$ in the
data compression problem, and the Bragg peaks at all harmonics of the
fundamental wave-number $q_{m_0}=2\pi m_0/d_0$ correspond to all harmonics of
the fundamental frequency $\omega_0=2\pi\beta m_0$ in the oscillatory
component of $R_n$. In other words, the parameter $\beta$ is conjugate, in
this sense, to $1/d_0$.

\section{Conclusion}

In this short paper, we have made an attempt to provide some insight into the
erratic behavior of the redundancy pattern of the Shannon code for lossless
data compression. The insight we propose is rooted in the physical point of
view, where the two modes of the behavior of the redundancy patterns are
respectively analogous to partial order and complete disorder of a wave diffraction
medium, which dictates the existence or non--existence of Bragg peaks
pertaining to perfectly constructive interference. It is hoped that this
physical insight contributes to the intuitive understanding of the redundancy
of the Shannon code and perhaps other codes as well.

Finally, we comment that the above analyses are, in principle, generalizable to
the finite--state Markov case (and indeed, Markov 
models have been proposed in
the diffraction setting too \cite{HT42},\cite{Welberry85}). 
When it comes to the Markov case, then both in the data compression problem
and in the HT model,
the role played by high powers of $C_m$ 
is essentially replaced by high powers of
state transition probability matrix whose entries are
weighted by the appropriate complex exponentials (which depend on $m$).
What matters then are the eigenvalues of this matrix.
More concretely, it is not difficult to see that 
the spectral radius, in both settings, never
exceeds unity. In the data compression problem, the critical behavior
is dictated by the existence or non--existence of
integer values $\{m\}$ for which the spectral radius is exactly
1. When such values of $m$ exist, then $R_n$ has an oscillatory behavior.
In the diffraction problem, the distinction between the two types of behavior
is dictated by the existence of values of $m$ for which one of the eigenvalues
is exactly equal to one.




\begin{thebibliography}{AA}

\bibitem{CD92}
R.~M.~Capocelli and A.~De Santis, ``On the redundancy of optimal codes
with limited word length,''
{\em IEEE Trans.~Inform.~Theory\/},
vol.~IT--38, no.~2, pp.~439--445, March 1992.

\bibitem{CL95}
P.~M.~Chaikin and T.~C.~Lubensky, {\it Principles of Condensed Matter
Physics}, Cambridge University Press, Cambridge England, 1995.

\bibitem{Gallager78}
R.~G.~Gallager, ``Variations on the theme by Huffman,'' 
{\em IEEE Trans.~Inform.~Theory\/},
vol.~IT--24, no.~6, pp.~668--674, November 1978.

\bibitem{GL88}
A.~Garg and D.~Levine, ``Speckle in the diffraction patterns of
Hendricks--Teller and icosahedral glass models,'' {\it Physical Review
Letters}, vol.\ 60, no.\ 21, pp.\ 2160--2163, 23 May 1988.

\bibitem{HT42}
S.~Hendricks and E.~Teller, ``X--ray interference in partially ordered layer
lattices,'' {\it The Journal of Chemical Physics}, vol.\ 10, no.\ 3, pp.\ 147--167,
March 1942.

\bibitem{Huffman52}
D.~A.~Huffman, ``A method for the construction of minimum-redundancy 
codes,'' {\it Proc.\ IRE}, vol.\ 40, pp.\ 1098--1101, 1952.

\bibitem{JS95}
P.~Jackuet and W.~Szpankowski, ``Asymptotic behavior of the Lempel--Ziv
parsing scheme and digital search trees,'' {\it Theoretical Computer Science},
vol.\ 144, pp.\ 161--197, 1995.

\bibitem{Krichevskii68}
R.~E.~Krichevskii, ``The relation between
redundancy coding and the reliability of information from a source,''
{\em Problems of Information Transmission (IPPI)\/}.
vol.~4, no.~3, pp.~37--45, 1968.

\bibitem{LS97}
G.~Louchard and W.~Szpankowski, ``Average redundancy of the
Lempel--Ziv code,'' {\em IEEE Trans.~Inform.~Theory\/},
vol.~43, no.~1, pp.~2--8, January 1997.

\bibitem{Rissanen86}
J.~Rissanen, ``Complexity of strings in the class of Markov sources,''
{\em IEEE Trans.~Inform.~Theory\/},
vol.~IT--32, no.~4, pp.~526--532, July 1986.

\bibitem{Savari98}
S.~A.~Savari, ``Variable--to--fixed length codes for predictable sources,''
{\it Proc.\ Data Compression Conference (DCC)}, Snowbird, UT, pp.\ 481--490,
1998.

\bibitem{SG97}
S.~A.~Savari and R.~G.~Gallager, ``Generalized Tunstall codes for sources with
memory,'' {\em IEEE Trans.~Inform.~Theory\/},
vol.~43, no.~2, pp.~658--668, March 1997.

\bibitem{Szp00}
W.~Szpankowski, ``Asymptotic average redundancy of Huffman (and other)
block codes,'' {\it IEEE Trans.\ Inform.\ Theory}, vol.\ 46, no.\ 7, pp.\
2434--2443, November 2000.

\bibitem{Welberry85}
T.~R.~Welberry, ``Diffuse X--ray scattering and models of disorder,''
{\it Rep.\ Prog.\ Phys.}, vol.\ 48, pp.\ 1543--1593, 1985.

\end{thebibliography}
\end{document}